\documentclass[aps, pre, twocolumn,amsmath, superscriptaddress,showkeys,showpacs]{revtex4-1}
\usepackage{graphicx}   
\usepackage{dsfont}
\usepackage{color}
\usepackage{epstopdf}
\usepackage{caption}
\usepackage{subcaption}
\usepackage{ragged2e}
\DeclareCaptionJustification{justified}{\justifying}
\begin{document}

\title{Coherent libration to coherent rotational dynamics via  chimeralike states and clustering in Josephson Junction array}
\author{Arindam Mishra}
\email{arindammishra@gmail.com}
\affiliation{Department of Physics, Jadavpur University, Jadavpur,  Kolkata  700032, India}

\affiliation{Center for Complex System Research Kolkata, Kolkata 700094, India}
 \author{Suman Saha}
 \email{ecesuman06@gmail.com}
  \affiliation{Department of Electronics, Asutosh College,  Kolkata  700026, India}
   \affiliation{Dumkal Institute of Engineering and Technology, Murshidabad 742406, India}
 \author{Chittaranjan Hens}
 \email{chittaranjanhens@gmail.com}
 \affiliation{Department of Mathematics, Bar-Ilan University, Ramat Gan 529002, Israel}
\author{Prodyot K. Roy}
\affiliation{Center for Complex System Research Kolkata, Kolkata 700094, India}
\affiliation{Department of Mathematics, Presidency University,  Kolkata 700073, India}

\author{Mridul Bose}
\affiliation{Department of Physics, Jadavpur University, Jadavpur,  Kolkata  700032, India}
\author{Patrick Louodop}
\affiliation{Instituto de F\'isica Te\'orica, Universidade Estadual Paulista, 
	01140-070 S\~ao Paulo, Brazil}
\affiliation{ 
	Department of Physics, University of Dschang, P.O. Box 67 Dschang, Cameroon}
\author{Hilda A. Cerdeira}, 
\affiliation{Instituto de F\'isica Te\'orica, Universidade Estadual Paulista, 
	01140-070 S\~ao Paulo, Brazil}
\author{Syamal K. Dana}
\affiliation{Center for Complex System Research Kolkata, Kolkata 700094, India}
\affiliation{CSIR-Indian Institute of Chemical Biology, Jadavpur,  Kolkata  700032, India}

\date{\today}

\begin{abstract}
An array of excitable Josephson junctions under global mean-field interaction and a common periodic forcing shows emergence of two important classes of coherent dynamics, librational and rotational motion in the weaker and stronger coupling limits, respectively, with transitions to chimeralike states and clustered states in the intermediate coupling range.   In this numerical study, we use the Kuramoto complex order parameter and introduce two measures, a libration index and a clustering index  to characterize the dynamical regimes and their transition and locate them in a parameter plane. 
 
\end{abstract}

                             
\maketitle
\par  
A surprising new phenomenon was reported in the last decade, namely, the chimera states \cite{Kuramoto, Strogatz, Martenes, Sen-2008, Omelchenko-chaos, Omelchenko-2013, Sheeba-2010, Sen-2013}  that emerge via a symmetry breaking of a homogeneous synchronous state in a large population of nonlocally coupled identical phase oscillators  into two coexisting spatially extended coherent and noncoherent subpopulations. Presently,
existence of chimera states has been reported in identical limit cycle oscillators \cite{Sen-2013, Davidsen},  chaotic systems \cite{Davidsen, Anna, Maistrenko, Kapitaniak, Schimdt} and very recently in excitable systems in presence of noise \cite{Semenova}. 
It drew special attention  after noticing a similar behavior in the brain of some sleeping animals  \cite{Kelso}. It has been now confirmed in laboratory experiments \cite{Tinsley, Murphy, Martenes-2013}. Most surprisingly, chimeralike states  were observed  in globally coupled network of identical oscillators \cite{Kaneko, Pikovsky, Sen-2014, Mishra, Hens} which was unexpected because of the presence of a perfect symmetry in such a network. The reason of symmetry breaking of a  homogeneous state into coexisting coherent and nocoherent states still remains a puzzle. 
\par In the meantime, more reports are coming on chimera states in many interesting systems, a network of neurons under different coupling forms \cite{Dibakar}, a Josephson junction array \cite{Lazarides} and chemical oscillators \cite{kiss}  under nonlocal coupling, which are of practical interest.  In particular, the Josephson junction, besides its main appeal as a  superconducting device, shows a rich variety of dynamics, excitability, bistability \cite{Dana, Levi, Strogatz1} and neuron-like spiking and bursting \cite{Dana, Dana1, Crotty, Lynch} that are of common interest in other areas of nonlinear science. In fact, synchronization as a symmetry preserving phenomenon in a globally coupled Josephson junction array \cite{Vlasov, Wissenfeld, Watanabe, Hilda} was studied, in the past, from the fundamental viewpoint of collective behaviors of oscillatory systems. It is now of general interest if symmetry breaking chimera states  emerge in a globally coupled Josephson junction array too.  
\par We report a search, in this paper, for chimera states in a Josephson junction array under global mean-field interaction if they exist at all and under what conditions?  
The existence of a state of order and turbulence was reported earlier \cite{Hilda} in a forced Josephson junction array under global mean-field influence, which showed signatures of chimera states, however, no categorical statement was made at that time. We revisit that parameter space of the Josephson junction array under the same condition and  confirm existence of chimeralike states. In the process, we notice two important classes of coherent states, one regular librational motion and a regular rotational motion in the array, which are typical dynamical features \cite{Strogatz1} of a single Josephson junction. In  cylindrical space  \cite{cylinder}, the trajectory of a junction is localized during a libration while it encircles the cylinder during a rotational motion (Fig. \ref{FIG.4}).  Most importantly, we observe a transition between the two coherent states for changing coupling interaction. Increasing the coupling strength from a weaker range, the coherent librational motion emerges  above a threshold and continues for a range of coupling, then transits to coherent rotational motion for large coupling via successive chimeralike states and clustered states in an intermediate coupling range.
In the chimeralike states, we notice coexistence of regular librational motion in a coherent subpopulation and chaotic rotational motion in another noncoherent subpopulation. In the clustered state, regular libration coexists with rotational motion in two subpopulations.  
\par We consider an array of identical Josephson junctions when each node of the network is driven by a radio-frequency (rf) signal.  We choose global mean-field interaction for the network and  identical parameters  as $\alpha = 0.2$ and $I = 0.021$ for all the junctions when an isolated junction remains in a stable steady state \cite{Dana}. The set of parameters are chosen almost identical to what was considered earlier \cite{Hilda}, for a revision of the  past result in search of chimera states. The rf forcing has  identical amplitude $I_{rf} = 0.595$ and frequency $\Omega_{rf} = 0.8$  for all the nodes that make them  oscillate periodically.
\par  The dynamics of the $i$-th node of the rf-forced junction array is described,
\begin{equation}
\dot{\phi_{i}} = y_{i}
\end{equation}
\begin{equation}
\dot{y_{i}} = I - \sin \phi_{i} - \alpha y_{i} + I_{rf}\sin (\Omega_{rf}t) + K\alpha Y
\end{equation}
where  $Y$= $\frac{1}{N} \sum\limits_{j=1}^N  {y_j}$ is the mean-field value of voltage $\dot\phi_i$=$y_i$ across all the junctions, $\alpha$=$[h/2 \pi e I R^2 C]^{1/2}$$= (\frac{1}{\beta})^{\frac{1}{2}}$ is the damping parameter, $\beta$ is the McCumber parameter, $I$ is the normalized constant bias current. 
$K$ defines the strength of mean-field interaction between the junctions.
 Increasing $K$ reveals various network dynamics and collective states,  two coherent states, chimeralike states, cluster states. 
\par To distinguish the emergent states and their dynamics,  we use  the complex Kuramoto order parameter ($r$) \cite{kuramoto-1} and introduce two new measures, a clustering index ($s$) and a libration index ($l$). 
The complex Kuramoto order parameter $r$ is,  
\begin{equation}
 re^{i\Phi} = \frac{1}{N}\sum_{j=1}^{N} e^{i\phi_{j}}
 \end{equation}
 where $\phi_j$ is the instantaneous phase of each junction $j$. When all the oscillators are coherent $r$=1 and  in an incoherent state, $r$=0 while $0<r<1$ implies partial synchronization or clustering. The chimera states belong to a class of  partial synchronization. 
 \par Since the order parameter $r$ cannot distinguish the chimera states from the cluster states for intermediate values of $0<r<1$, we introduce a clustering index $s$, 
 \begin{equation}
 s = \frac{max(n)}{N}u
 \end{equation}
 where $u = 1 - \Theta(\delta_1 - d)$ and $d = max(n) -\langle n\rangle$ and $\Theta(.)$ is the Heaviside step function and $\delta_1$ is an arbitrary small number,  $n(t)$ is the number of distinct states counted (using standard numerical routine) at every instant of time $t$ in the time evolution of the network, $\langle n\rangle$ denotes the average in a long run. The $max(n)$ is the largest possible value of $n$. A clustered state (single or multiple) is now clearly distinguished by $s = 0$. It excludes a   cluster state when $0<s<1$ but detects  existence of chimeralike states if $0<r<1$.
\par Next  the libration index $l$ is introduced basically to characterize the dynamical features of the junctions in different collective states,  
\begin{equation}
l = \frac{1}{N} \sum_{j=1}^{N} \Theta_{j}
\end{equation}
 with $\Theta_{j} = \Theta(\delta_2 - m_{j})$ where $\delta_2$ is another arbitrarily chosen small threshold, $\Theta(.)$ is defined above and $m_{j}$ is
\begin{equation}
m_{j} = 1 - 0.5[max(\cos (\phi_{j}(t))) - min(\cos (\phi_{j}(t)))]
\end{equation}
 To determine $m_j$  for the $j$th oscillator, we calculate cos($\phi_{j}(t)$) for all instant of time, which varies from $0$ to $2\pi$ for rotational motion when $m_{j} = 0$. 
In libration, since the trajectory of an oscillator never crosses the $\phi = \pi$ line, $m_{j}$ is a positive number.
Finally it determines   $l = 0$ for oscillators in libration and  $l = 1$ when they are in rotational motion. A value of $0<l<1$ indicates  coexistence of librational and rotational motion in subpopulations of the junctions; see the Supplemental Material (SM) for details \cite{SM}. 
\begin{figure}[h!]
	\centering
	\hspace{-30pt}
	\includegraphics[width = 8cm,height = 5cm]{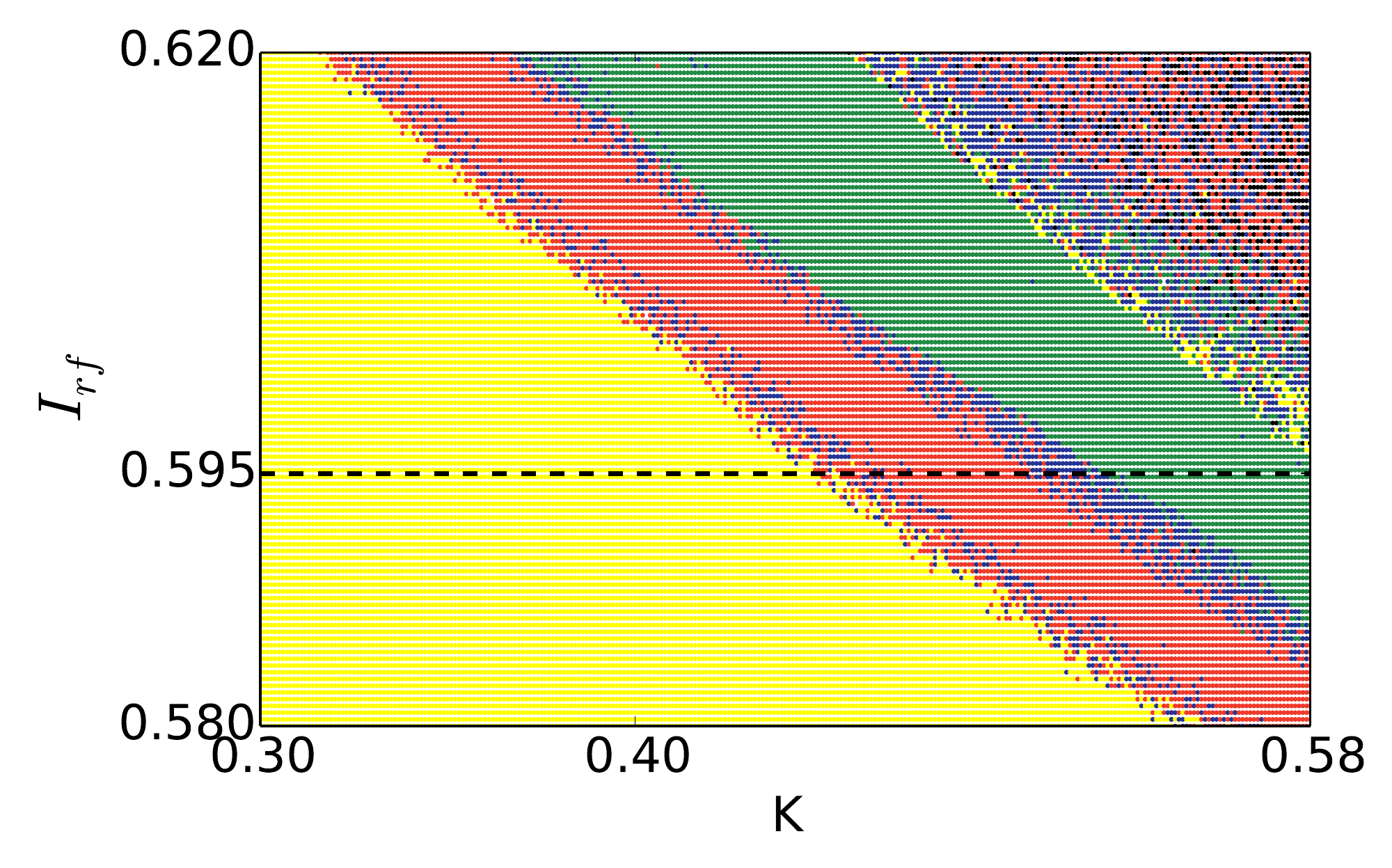}	
	\caption{(color online)  Different dynamical states in $K - I_{rf}$ space for $\alpha = 0.2$, $I = 0.021$ and $\Omega_{rf} = 0.8$. Yellow and green  regions represent coherent librational  and coherent rotational motion,  respectively and, blue and red regions denote cluster and chimera states, respectively. Black dots (upper corner) represent desynchronized states. We perform our numerical study in Fig.~2 along the horizontal dashed line,  $I_{rf} = 0.595$. }
	\label{FIG.1}
\end{figure}	

\par Figure \ref{FIG.1} shows distinct  dynamical regimes in the $K - I_{rf}$ space where each point is plotted in color using a combination of all the three above measures. The regions of coherent libration and coherent rotation are denoted by yellow and green colors, respectively. The red color represents chimeralike states where coherent oscillators are in libration and incoherent oscillators in rotational motion. The cluster state is depicted by blue color where a mixed population with libration and rotation exists. Black dots represent desynchronized states. A region of messy colors is seen on the top right corner where cluster and chimera states and even coherent states coexist, which is not the focus of our current interest. As a specific example, we vary $K$ along the horizontal dashed line ($I_{rf} = 0.595$) shown in Fig.\ref{FIG.1} and follow a transition from a coherent librational state to another coherent rotational state through the intermediate chimeralike and cluster states as mentioned above. 
\begin{figure}[h!]
	\centering
	\centering
	\includegraphics[width = 9cm,height = 7cm]{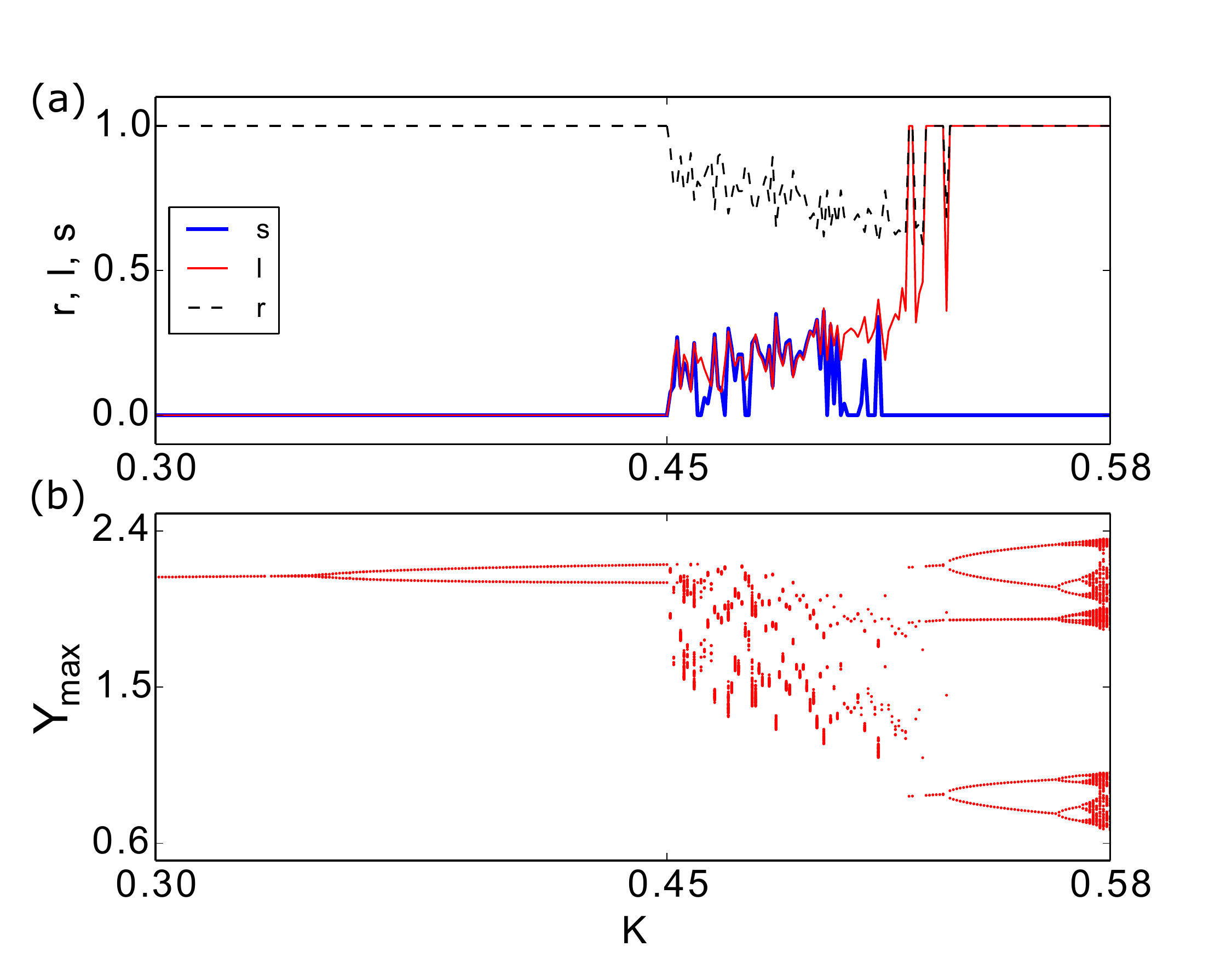}	
	\centering	
	\caption{(Color online) Plots of $r$ (dashed line), $l$ (red line) and $s$ (blue line) with coupling strength $K$ in (a). A bifurcation plot of $Y_{max}$ against $K$ in (b). $\alpha=0.2$, $I = 0.021$, $\Omega_{rf}=0.8$, $I_{rf} = 0.595$.}
	\label{FIG.2}
\end{figure}
\begin{figure}[h!]
	\centering
	\hspace{-20pt}
	\begin{subfigure}[b]{0.5\columnwidth}
		\centering
		\includegraphics[width = 4cm,height = 4cm]{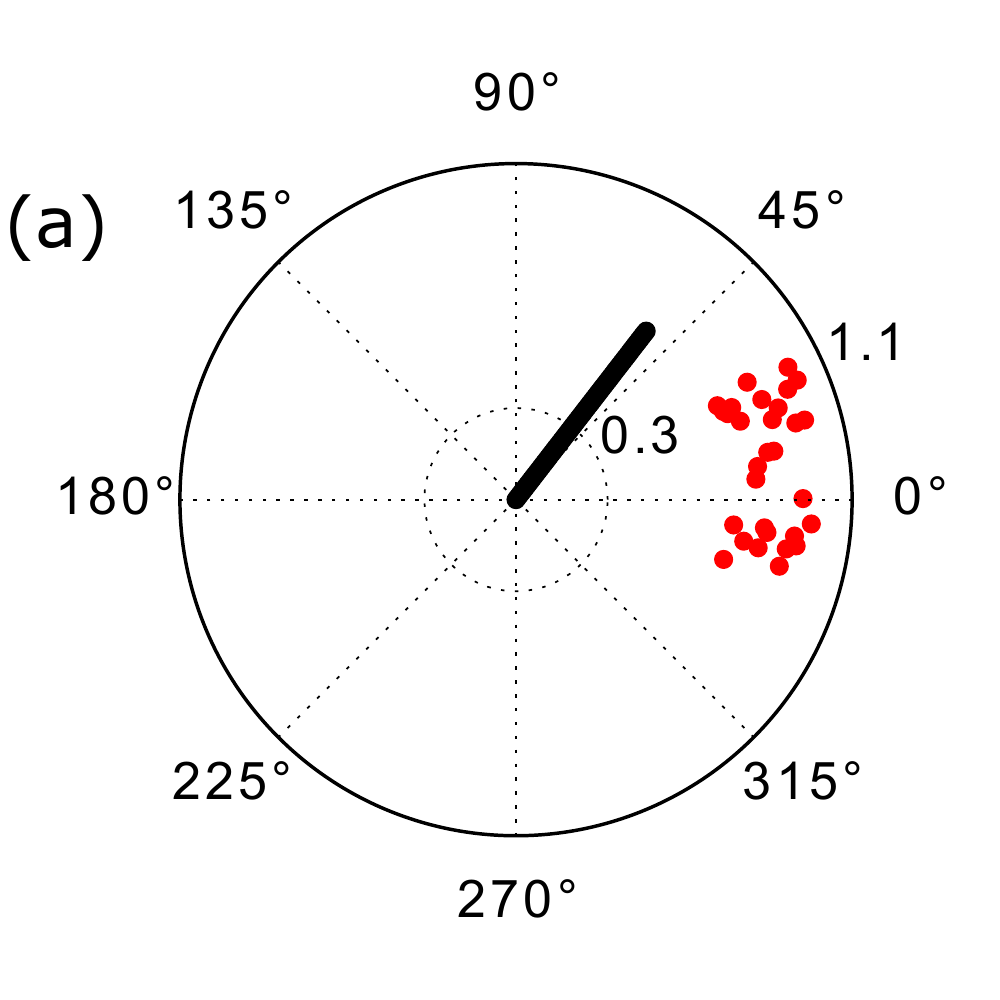}
	\label{FIG.3(a)}
	\end{subfigure}
	\hspace{-15pt}
	\begin{subfigure}[b]{0.5\columnwidth}
		\centering
		\includegraphics[width = 5.25cm,height = 3.25cm]{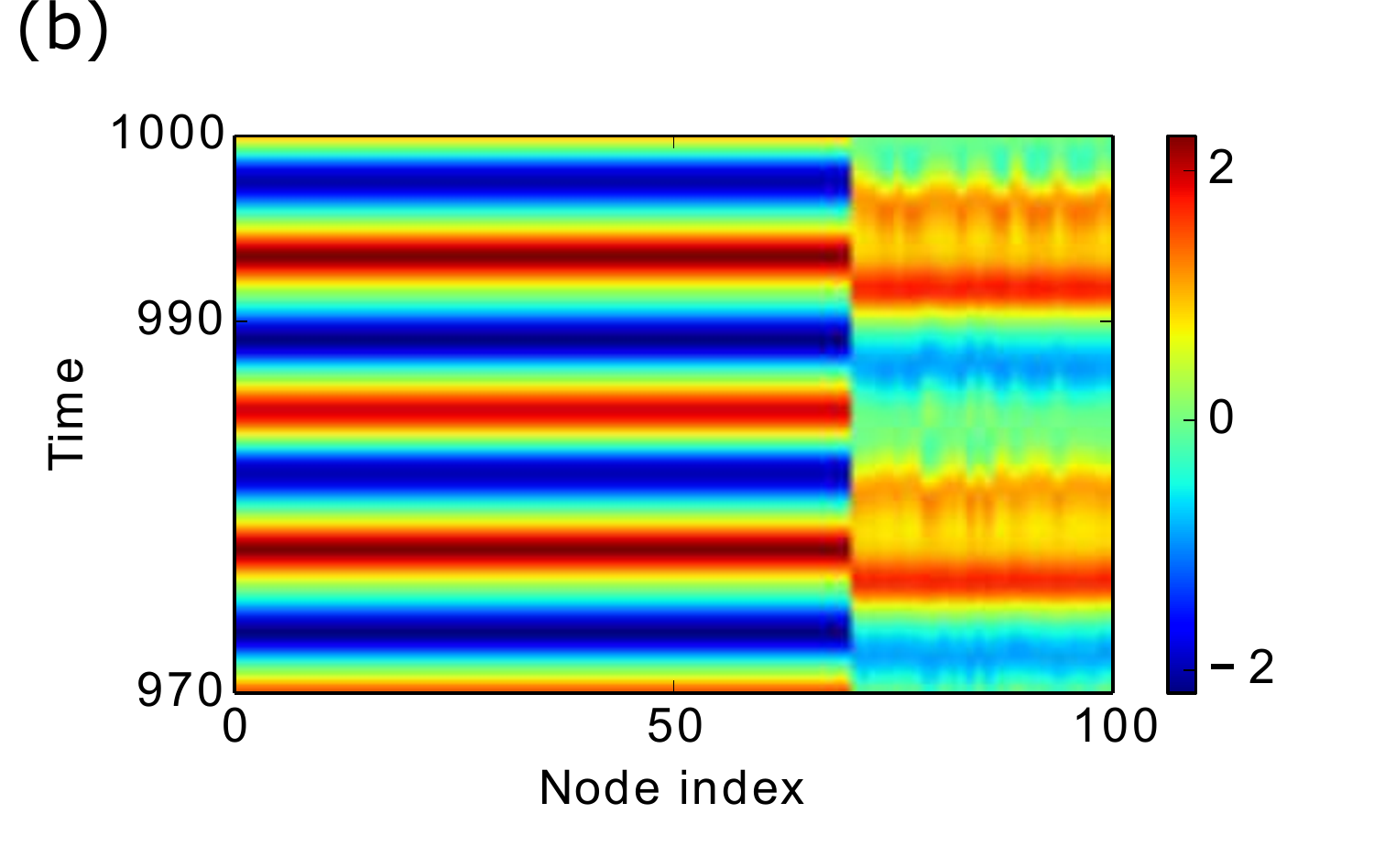}
		\label{FIG.3(b)}
	\end{subfigure}	
	\caption{(Color online) Snapshot of phases of all junctions in a polar plot (a), their spatiotemporal dynamics (b) confirm chimeralike states ($K=0.49$).  Red circles, black line represent incoherent and coherent subpopulations, respectively in (b).}
	\label{FIG.3}
\end{figure}\\

\par Before describing further the collective dynamics, we reduce the coupled Eqs.(1)-(2) to their averaged equations (equivalent to the motion of the center of mass of our system), 
\begin{equation}
\dot{X} = Y
\end{equation}
\vspace{-20pt}
\begin{equation}
\dot{Y} = I - \frac{1}{N} \sum_{j=1}^{N} \sin \phi_{j} - (1 - K)\alpha Y + I_{rf}\sin (\Omega_{rf}t)
\end{equation}
where $X = \frac{1}{N} \sum_{j=1}^{N} \phi_{j}$ and $Y = \frac{1}{N} \sum_{j=1}^{N} y_{j}$ define mean phase and mean voltage of the junction array. We simultaneously simulate the averaged equations and the original coupled equations. The average of $\langle \sin \phi_i\rangle$ is simulated from the original coupled equations and substituted into the averaged equation.  

\par Figure \ref{FIG.2}(a) plots $r$ (dashed line), $s$ (blue line) and $l$ (red line) for varying $K$ in a range of $0.3$ to $0.58$. 
As we increase $K$, $r$ becomes 1  and $s=0$ (i.e., coherent state or single clustered state) above a threshold (not shown here) and continues for a range of coupling   
 until $r$ starts decreasing at a critical value, $K=0.451$ when $Y$ becomes chaotic. In the latter range of $K$ values, both $r$ and $s$ fluctuates, $0<r<1$ and $0<s<1$, where chimeralike states emerge as a partial synchronization state. For further increase of $K$ above another critical value, $K=0.512$, $r$ starts fluctuating between 1 and intermediate values $0<r<1$ intermittently in small windows of $K$, which signifies a switching between a single cluster and multicluster states for small changes of $K$ values until it reaches $K=0.532$. However, $s=0$ all along for $K>0.512$ confirms the presence of   single  or multicluster states. In fact, the single cluster or the coherent state emerges at $K=0.532$ when $r=1$ ($s=0$). The chimeralike states emerge only in the range of  $K=0.451-0.512$ where $r$ shows a decreasing trend and $s$ shows a reverse trend  except in the clustered states (small windows of single clusters and multicluster states).
 \begin{figure}[h!]
 	\centering
 	\hspace{-15pt}
 	\begin{subfigure}[b]{0.5\columnwidth}
 		\centering
 		\includegraphics[width = 4cm,height = 4cm]{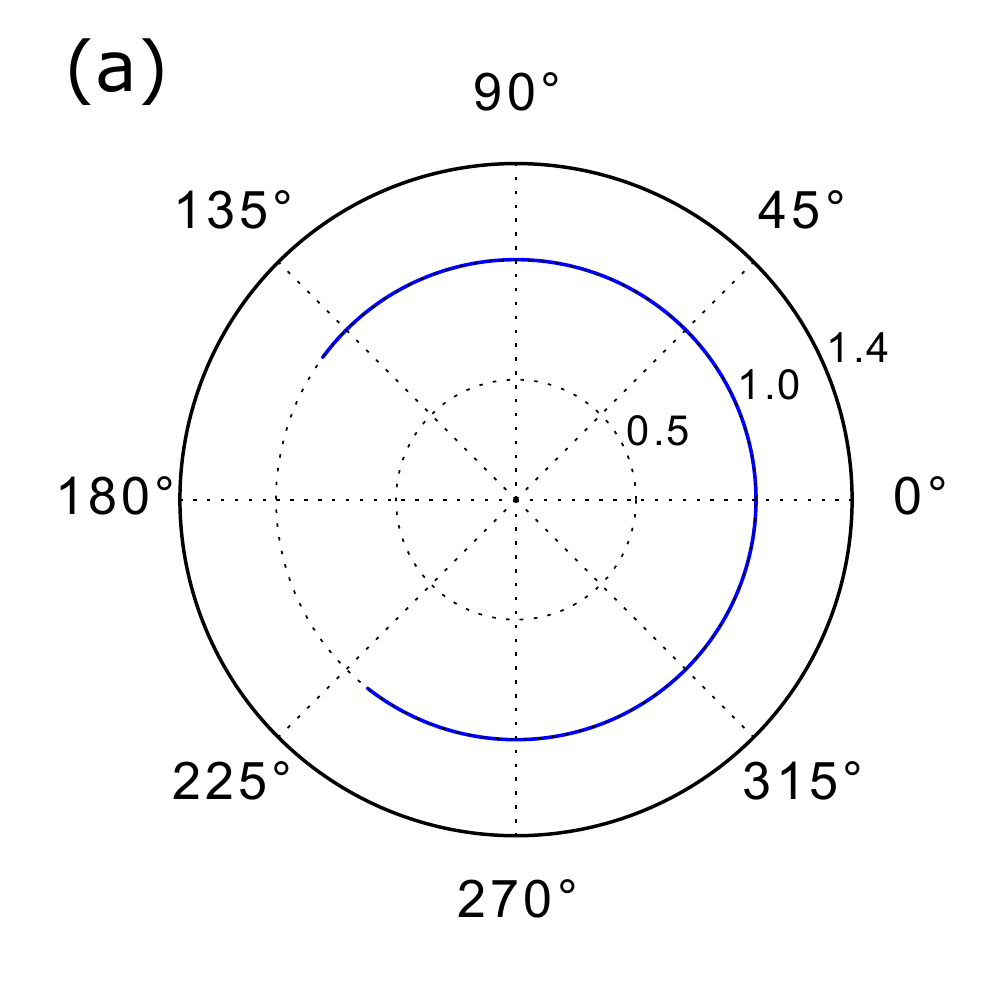}
 		\label{FIG.4(a)}	
 	\end{subfigure}
 	\hspace{-10pt}
 	\begin{subfigure}[b]{0.5\columnwidth}
 		\centering
 		\includegraphics[width = 3cm,height = 4cm]{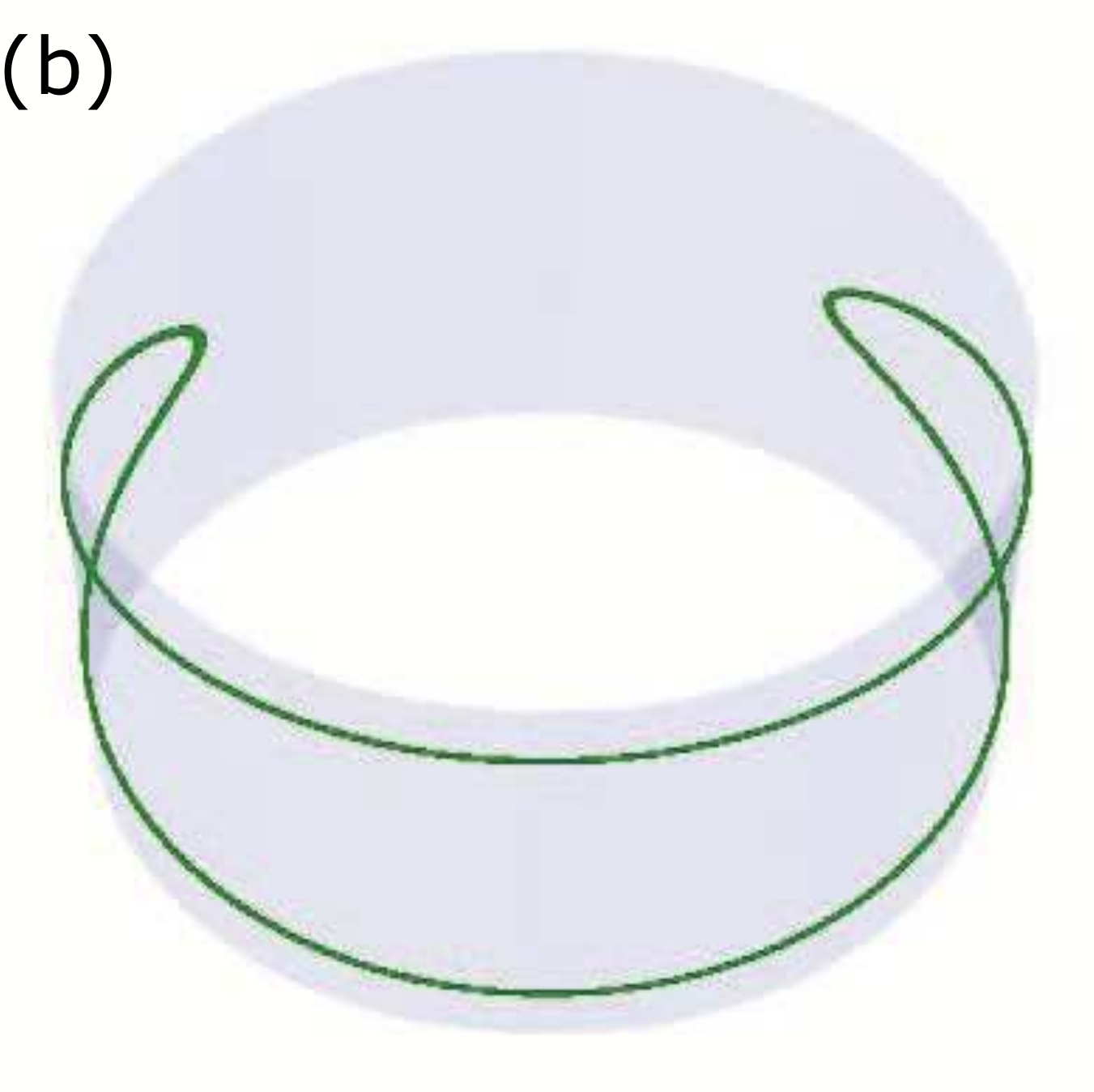}
 		\label{FIG.4(b)}	
 	\end{subfigure}\\
 	\hspace{-15pt}
 	\begin{subfigure}[b]{0.5\columnwidth}
 		\centering
 		\includegraphics[width = 4cm,height = 4cm]{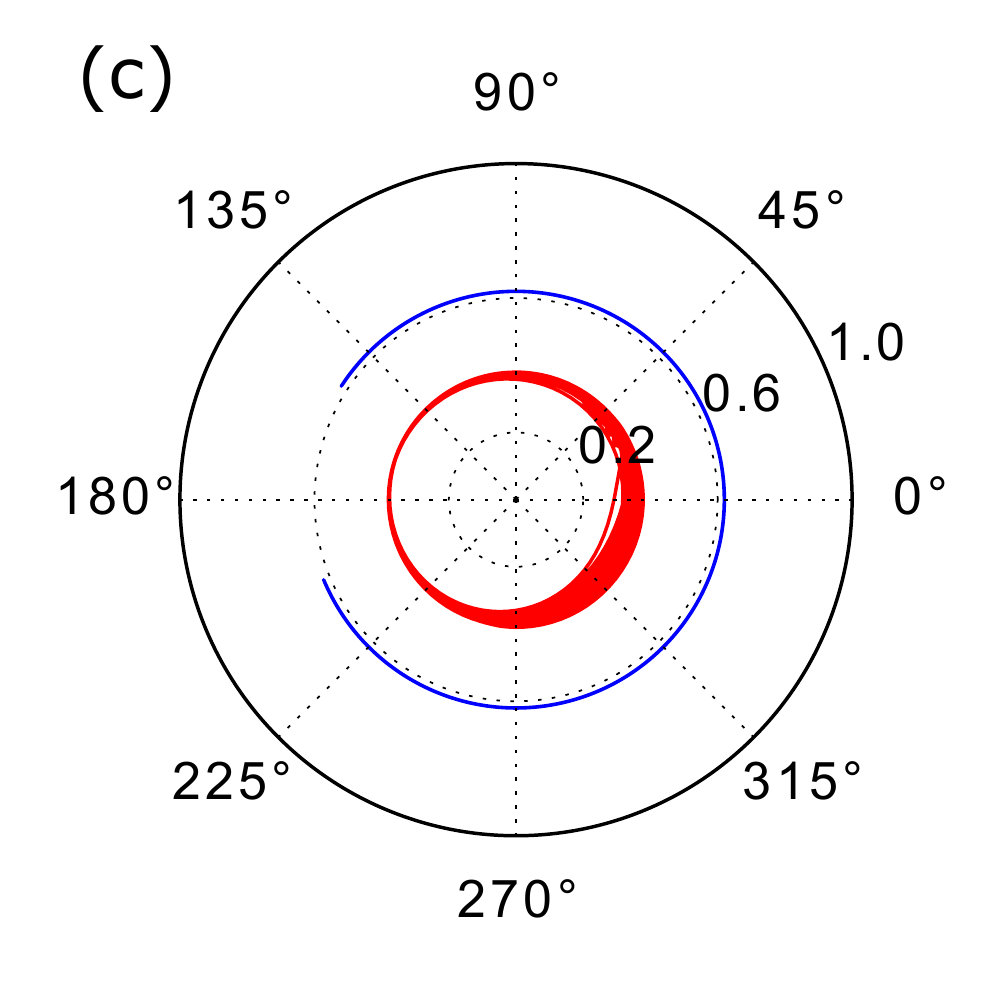}
 		\label{FIG.4(c)}	
 	\end{subfigure}
 	\hspace{-10pt}
 	\begin{subfigure}[b]{0.5\columnwidth}
 		\centering
 		\includegraphics[width = 3cm,height = 4cm]{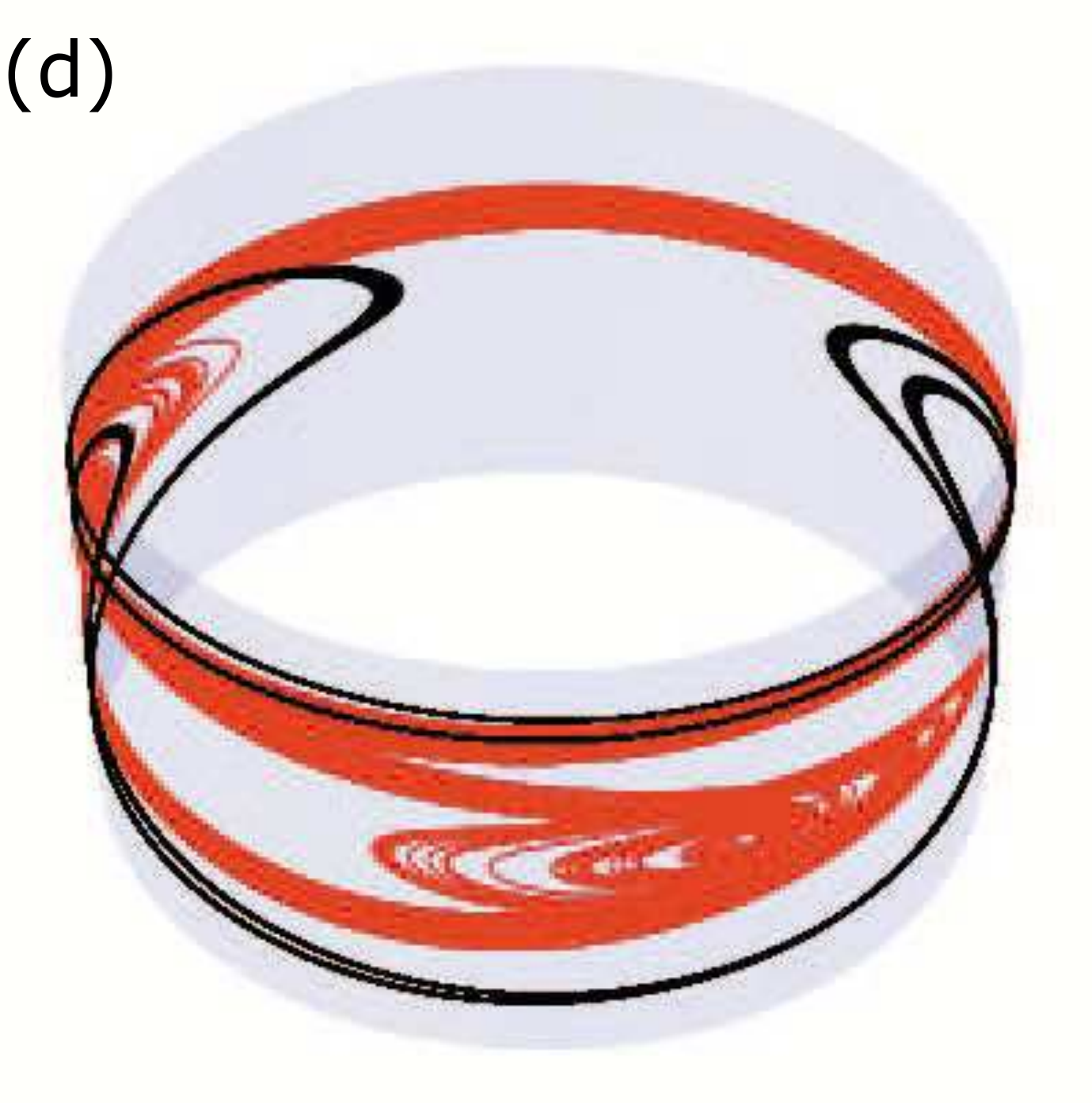}
 		\label{FIG.4(d)}	
 	\end{subfigure}\\
 	\hspace{-15pt}
 	\begin{subfigure}[b]{0.5\columnwidth}
 		\centering
 		\includegraphics[width = 4cm,height = 4cm]{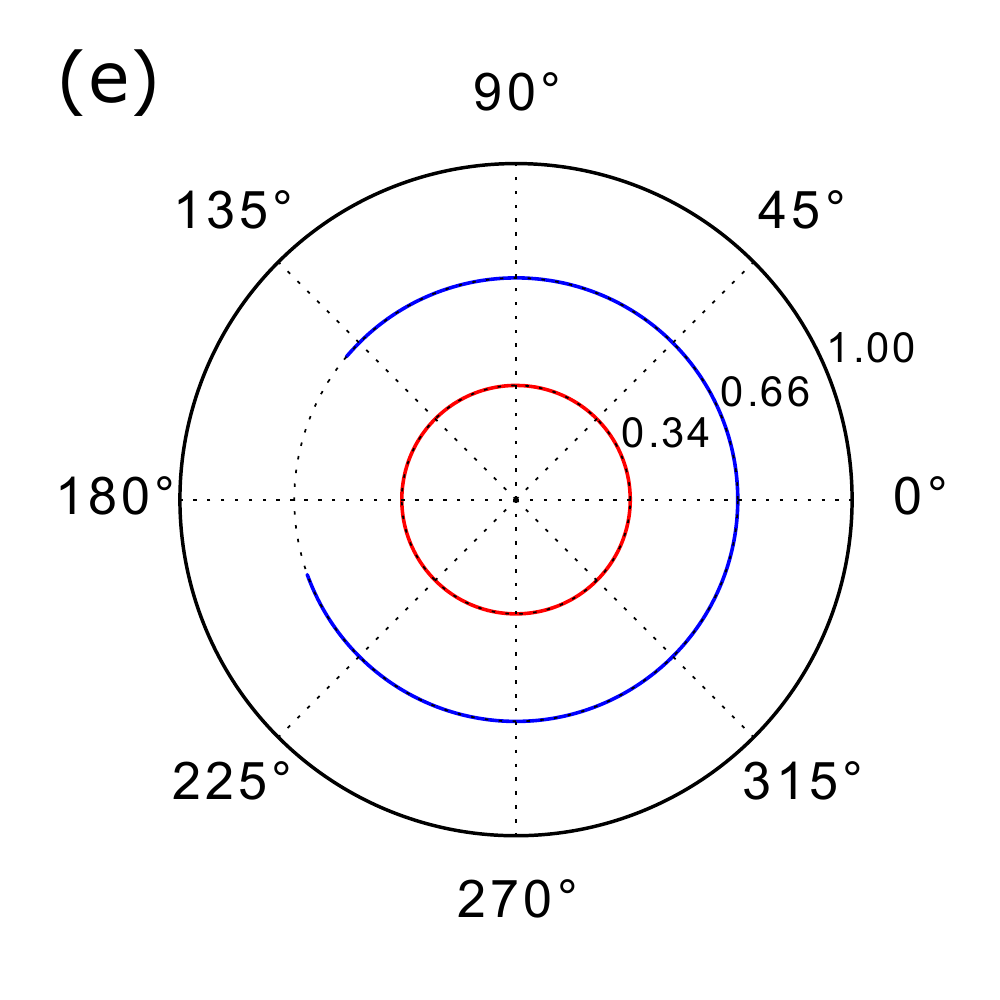}
 		\label{FIG.4(e)}	
 	\end{subfigure}
 	\hspace{-10pt}
 	\begin{subfigure}[b]{0.5\columnwidth}
 		\centering
 		\includegraphics[width = 3cm,height = 4cm]{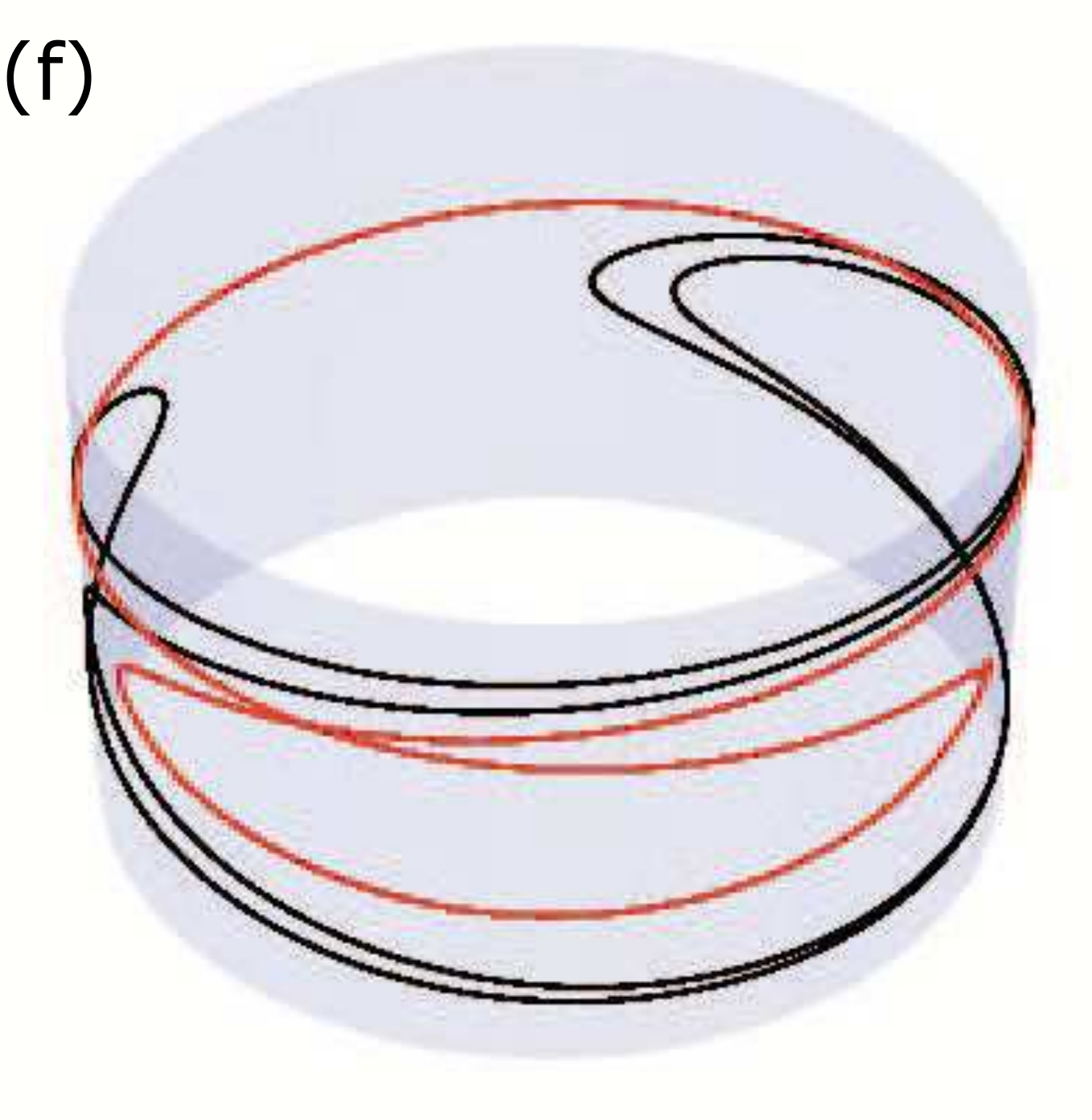}
 		\label{FIG.4(f)}	
 	\end{subfigure}\\
 	\hspace{-15pt}
 	\begin{subfigure}[b]{0.5\columnwidth}
 		\centering
 		\includegraphics[width = 4cm,height = 4cm]{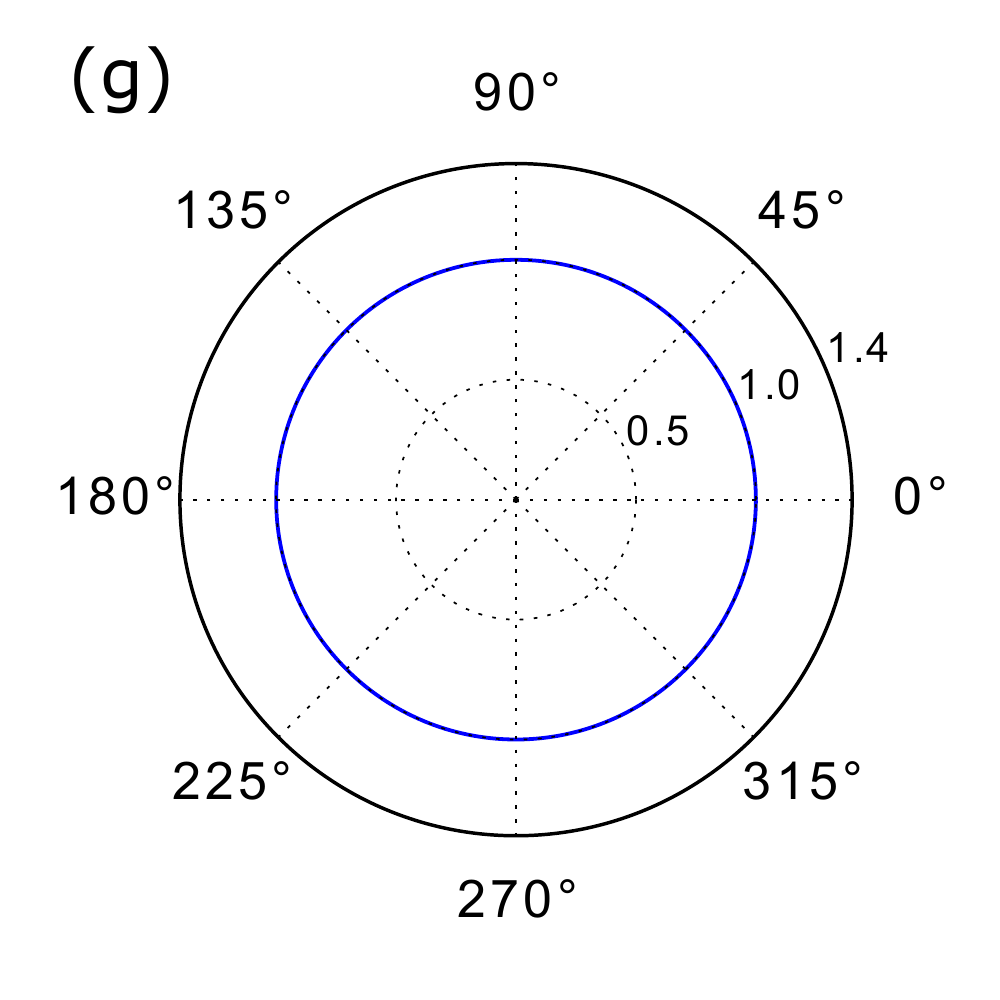}
 		\label{FIG.4(g)}	
 	\end{subfigure}
 	\hspace{-10pt}
 	\begin{subfigure}[b]{0.5\columnwidth}
 		\centering
 		\includegraphics[width = 3cm,height = 4cm]{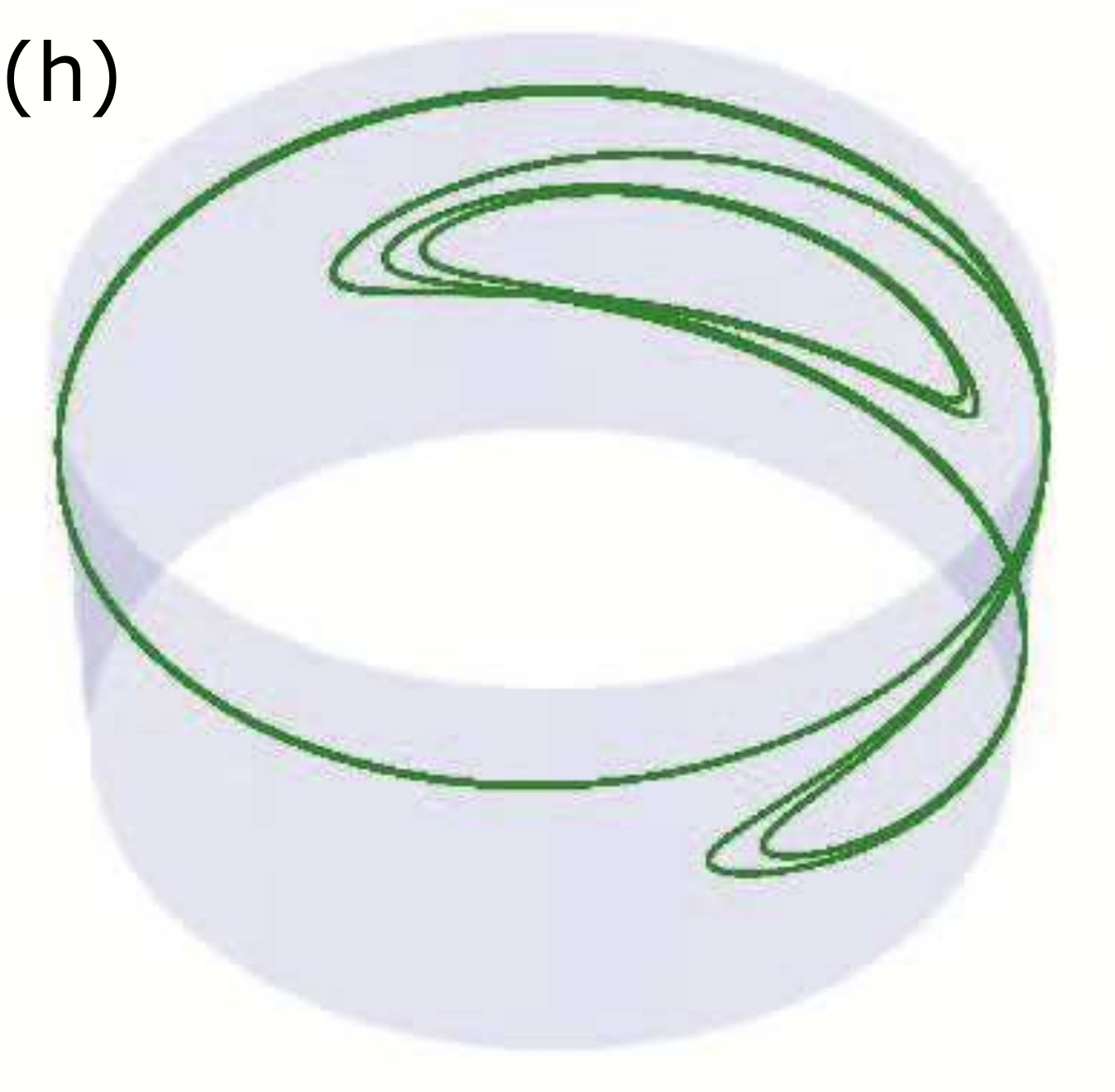}
 		\label{FIG.4(h)}
 	\end{subfigure}	
 	\caption{(Color online) $r-\Phi$ plot (left column) and phase portrait ($y$ vs. $\phi$) in cylindrical space (right column). Coherent libration (a,e) for $K=0.305$, chimeralike states (b,f) for $K=0.49$,  cluster states (c,g) of coexisting libration and rotational motion for $K=0.523$, coherent  rotational motion (d,h) for $K=0.56$.}
 	\label{FIG.4}
 \end{figure} 
\par Figure \ref{FIG.2}(b) presents a bifurcation diagram of $Y_{max}$ with $K$. It reveals that the collective dynamics in the left coherent region (cf. upper panel) bifurcates from a single period to period-2 and larger periods, and followed by chaos in the chimera region.  In this chimera region, the $Y_{max}$ plot  indicates chaotic behavior where the average  $Y$ was taken on  two subpopulations, one in coherent periodic motion and another in non-coherent chaotic motion. A small window of  multiclustered state (period-3) exists immediately after the chimera region followed by  single cluster higher periodic rotational orbits (period-6) on the right (lower panel) that again bifurcates via period-doubling to chaos, however, remains in coherent rotational state. Note that $r, s$ and $l$ do  not fluctuate here that makes its distinction from the chaotic chimera states. 
\par For a demonstration of the chimeralike states, we present  ($K=0.49$)  a snapshot of phases of all the junctions in a polar plane in Fig. \ref{FIG.3}(a). The incoherent subpopulation is clear from the distribution of phases of individual junctions (red circles) and the coherent junctions are aligned along the black line. The spatiotemporal dynamics of the voltage variable ($y$) of all the junctions is plotted in Fig. \ref{FIG.3}(b) for a long run that further confirms the existence of the chimeralike states, coexisting coherent and incoherent subpopulations. 

\par Figure \ref{FIG.4} describes the dynamical characters in different collective states (two coherent states, clustered state and chimeralike states) of the junctions in $r-\Phi$ plots (left column) and their phase portraits in cylindrical space (right column).
In the lower range of $K<0.451$ (cf. Fig. \ref{FIG.2}), the coherent or single cluster dynamics of the junctions is librational ($l=0$, $s=0$)  when the $r-\Phi$ plot  in Fig.~4(a) shows an incomplete rotation (blue line, $r=1$) and it is confirmed by its trajectory (green line) in a cylindrical space in Fig.~\ref{FIG.4}(b). In contrast, the coherent or single cluster dynamics of the junctions at the other end (cf. Fig. \ref{FIG.2}) for larger $K>0.532$ is complete rotational ($l=1$, $s=0$)  as shown  in the $r-\Phi$ plot (blue line, $r=1$) and its trajectory (green line) in cylindrical space in Figs.~\ref{FIG.4}(g) and \ref{FIG.4}(h), respectively. In the intermediate range, $0.451<K<0.512$,  $l$ fluctuates ($0<l<1$) and almost exactly follows the $r$ fluctuation ($0<r<1$) where $r$ shows a decreasing trend (Fig. \ref{FIG.2}) and we observe the chimeralike states. In the chimeralike states, the coherent subpopulation is in librational motion  (blue line, $r \approx 0.6<$1) and the incoherent subpopulation (red lines) coexists in rotational motion (red line, $r<1$; around 0.2) as shown in the $r-\Phi$ plot in Fig.~\ref{FIG.4}(c). In cylindrical space in Fig.~\ref{FIG.4}(d)), the trajectories of the coherent subpopulation (black line) confirm their regular librational motion and  the incoherent (red lines) counterpart in chaotic rotational motion. In the range of $K=0.512-0.532$, as mentioned earlier, the clustered states (cf., $l$=0 in Fig. \ref{FIG.2}) are seen where both the $r-\Phi$ plot and the trajectory in cylindrical space in Figs.~\ref{FIG.4}(e) and \ref{FIG.4}(f), respectively, confirm existence of coexisting subpopulations in regular rotational and regular librational motion.

In summary, we revisited an earlier study \cite{Hilda} on  collective dynamics of a globally coupled Josephson junction array under a common rf forcing where  order and turbulent states were reported to coexist although no categorical statement about the existence of chimeralike states was made there.  Our present numerical study confirmed that chimeralike states  indeed existed there. Furthermore, we explored two important classes of coherent states, a librational motion and a rotational motion, and an interesting process of transition from one to the other via successive emergence of chimeralike states and cluster states when the coupling strength was increased. This phenomenon of nontrivial transition  was not limited to a particular set of parameters used earlier but existed in a broader parameter range (a second example is presented in the Supplementary material \cite{SM}). The variety of dynamics, libration and rotational motion, in the junction array and their collective states were identified, in parameter space, using the  Kuramoto order parameter($r$) and  by introducing two new measures, a librational index ($l$) and a clustering index ($s$) and illustrated them in a cylindrical space.
 \par S. K. D. and P. K. R. acknowledge support by the CSIR (India) under the Emeritus Scientist Scheme. A. Mishra is supported by the UGC India. H. A. C. and S. K. D. thank  ICTP-SAIFR and FAPESP grant 2011/11973- 4 for partial support. P. L. acknowledges support by the FAPESP grant 2014/13272-1.
  S. S.  acknowledges support by the DST (India) and C. H. is supported by the PBC, Israel.


\begin{thebibliography}{10}
\bibitem{cylinder} The natural phase space of the trajectory of a Josephson junction is cylindrical. We wrap up the phase space on a cylinder. In this phase space, the periodic behavior of phase is more explicit. The radius of the cylinder ($\rho$) is taken as unity and its relation to the Cartesian coordinate is given by, $x_{1} = \rho \cos \phi$, $x_{2} = \rho \sin \phi$ and $x_{3} = y$.
\bibitem{initial1} For the network of RCSJs, we construct a general 'V'-shaped initial states for $y_{i}$ between 'range1' and 'range2' as $y_{i0}$ = range1 + (range2$-$range1)$(2\frac{i}{N})$ for $1 \leq i \leq \frac{N}{2}$ and $y_{i0}$ = range1 $-$ (range2$-$range1)$(2\frac{i}{N}-2)$ for $\frac{N}{2} + 1 \leq i \leq N$  with added small random fluctuations. All initial states for $\phi$-variables are set at zero. For our simulation we choose range1 $= 2.0$ and range2 $= -2.0$.
\end{thebibliography}
\end{document}


\beginsupplement
\title{Supplementary Material:\\ Coherent libration to coherent rotational dynamics via  chimeralike states and clustering in Josephson Junction array}
\author{Arindam Mishra}
\email{arindammishra@gmail.com}
\affiliation{Department of Physics, Jadavpur University, Jadavpur,  Kolkata  700032, India}

\affiliation{Center for Complex System Research Kolkata, Kolkata 700094, India}
\author{Suman Saha}
\affiliation{Department of Electronics, Asutosh College,  Kolkata  700026, India}
\affiliation{Dumkal Institute of Engineering and Technology, Murshidabad 742406, India}
\author{Chittaranjan Hens}
\affiliation{Department of Mathematics, Bar-Ilan University, Ramat Gan 529002, Israel}
\author{Prodyot K. Roy}
\affiliation{Center for Complex System Research Kolkata, Kolkata 700094, India}
\affiliation{Department of Mathematics, Presidency University,  Kolkata 700073, India}

\author{Mridul Bose}
\affiliation{Department of Physics, Jadavpur University, Jadavpur,  Kolkata  700032, India}
\author{Patrick Louodop}
\affiliation{Instituto de F\'isica Te\'orica, Universidade Estadual Paulista, 
	01140-070 S\~ao Paulo, Brazil}
\affiliation{ Laboratory of Electronics, Faculty of Science, 
	Department of Physics, University of Dschang, P.O. Box 67 Dschang, Cameroon}
\author{Hilda A. Cerdeira}, 
\affiliation{Instituto de F\'isica Te\'orica, Universidade Estadual Paulista, 
	01140-070 S\~ao Paulo, Brazil}
\author{Syamal K. Dana}
\affiliation{CSIR-Indian Institute of Chemical Biology, Jadavpur,  Kolkata  700032, India}
\affiliation{Center for Complex System Research Kolkata, Kolkata 700094, India}
\date{\today}
\maketitle
In this Supplementary material we describe, in more detail, how to estimate the new indices, namely, the clustering index ($s$) and the libration index ($l$) to characterize various collective dynamics of the junction arrays for increasing mean-field interaction ($K$). Moreover, we present a second example (Fig.~S1) with a different parameter region (cf. Fig.~2, upper panel, in the main text) where we observe a similar transition from libration to rotation via chimeralike states and cluster states with increasing $K$. It basically reveals that our observation is true for a reasonably broad parameter values, especially for a range of $\alpha$ values.
 \par The Kuramoto order parameter tells us about emergence of a coherent state or synchronization ($r = 1$), complete disordered state or incoherent state ($r = 0$ in the thermodynamic limit) and about the partially synchronized states ($0<r<1$) with varying coupling interaction between an ensemble of oscillators. However, it can not distinguish between the chimera and cluster states when the value of the order parameter could be $0<r<1$ for  both the states. It tells us nothing about the dynamics of the system. In our study, we distinguish different collective behaviors by defining two new measures namely, a libration index ($l$) and cluster index ($s$) besides the complex Kuramoto order parameter($r$). The libration index ($l$) especially distinguish  the collective dynamics  of the ensemble of the junctions, the libration motion and the rotational motion.
 
 \subsection{Clustering index ($s$)} We define the clustering index ($s$) to distinguish chimera states from the cluster states. In  cluster states, oscillators assemble themselves in one or more groups and the number of groups does not change with time. In chimera states,  the number of groups changes with time as the incoherent subpopulation oscillates randomly. At first we numerically calculate how many different groups are there at every instant of time and call it as $n(t)$ and then calculate $d = max(n) - \langle n\rangle$ where $\langle .\rangle$ denotes the time average for a long run. Since in cluster states the maximum number groups of oscillators or clusters remain unchanged, the time average and the maximum number of groups or clusters are same in time and hence we obtain $d = 0$ in cluster states. While the groups change with time in chimera states due to randomly oscillating incoherent population and thus $d > 0$ for chimera states. Once again we encounter numerical errors coming from the fact that, at a certain instant of time, two clusters may come very close to each other and we can not distinguish them as separate states although it rarely happens. As a result we get an estimate of $d$-value as a small positive number rather than zero in  a cluster state. To get rid of this, we use the  Heaviside function once again and define $u$ as $u = 1 - \Theta(\delta_1 - d)$  where $\delta_1$ is assumed to be an arbitrarily small quantity. For cluster states $u = 0$ and for chimera state $u =1$. 
 Finally we rescale it and define $s$ as
 \begin{equation}
 	s = \frac{max(n)}{N}u
 \end{equation}
 For cluster state $s = 0$, for fully desynchronized state $s = 1$ (as $max(n)$ is equal to the number of oscillators $N$ in this state) and it takes any intermediate value for chimera state. 
 \begin{table}[h!]
 	\centering
 	\caption{Values of $r, l, s$ for different dynamical states.}
 	\begin{tabular}{ |p{2cm}|p{1.75cm}|p{1.75cm}|p{1.75cm}|  }
 		\hline
 		Dynamical States & $r$ & $s$ & $l$\\
 		\hline
 		coherent libration & 1 & 0& 0\\
 		\hline
 		coherent rotation & 1  & 0 &1\\
 		\hline
 		cluster states &$0<r<1$ & 0& $0<l<1$\\
 		\hline
 		chimera states &$0<r<1$ & $0<s<1$& $0<l<1$\\
 		\hline
 	\end{tabular}
 	
 	\label{table:1}
 \end{table}
 
  \subsection{Libration index ($l$)} 
  In librational motion, the phase of an oscillator does not cover the whole $[0,2\pi]$ space (similar to a conventional pendulum). But in rotation it covers the whole $[0,2\pi]$ space (similar to an inverted  pendulum). We calculate cosine value of $\phi_{j}(t)$ (where $j$ is the oscillator index) for all instants of time. If the oscillator is in rotational motion, it rotates completely from $0$ to $2\pi$. So, maximum and minimum values of $\cos (\phi_{j}(t))$ will be $1$ for $\phi_{j} = 0$ and $-1$ for $\phi_{j} = \pi$. But it is not true for libration. For libration motion, $max(\cos (\phi_{j}(t))) = 1$ but $min(\cos (\phi_{j}(t)))$ will not be $-1$ since the oscillator never crosses $\phi = \pi$ line during such motion.
  Then we calculate 
  \begin{equation}
  m_{j} = 1 - 0.5[max(\cos (\phi_{j}(t))) - min(\cos (\phi_{j}(t)))]
  \end{equation}
  It is clear that for an oscillator in rotational motion, $m_{j} = 0$ and a positive number for an oscillator in librational motion. However, when we calculate numerically the value of $m_{j}$ it never be exactly zero for rotational motion. (It depends upon the time step size of numerical integration. The more the step size is small, the more it approaches to $0$). That is why we use the Heaviside step function to overcome the numerical error as $\Theta_{j} = \Theta(\delta_2 - m_{j})$ where $\delta_2$ is assumed again as an arbitrary small number (for our study we take time step size as 0.01 and $\delta = 0.05$). So $\Theta_{j} = 0$ for an oscillator in libration and $\Theta_{j} = 1$ for rotation. And finally we define libration index ($l$) by taking the average, 
  \begin{equation}
  l = \frac{1}{N} \sum_{j=1}^{N} \Theta_{j}
  \end{equation}
  If all oscillators are in libration, then $l = 0$ and $l = 1$ if all are in rotation. Any intermediate value defines a mixed state that indicates a fraction of oscillators coexisting in rotational motion.
\subsection{Transition from libration to rotational motion}
The transition from coherent libration to coherent rotation via chimera and cluster states is shown to emerge in the junction arrays for a choice of different set of parameters in FIG.~S1 (cf. Fig.~2, upper panel, in the main text). 
We show an example of another set of parameters ($\alpha = 0.25$ and $I_{rf} = 0.6$; cf. Fig~2, where  $\alpha = 0.2$, $I_{rf}=0.595$), and observe a  similar kind of transition. The critical values of transitions and the range of $K$ values for various dynamics only change. The coupling window (all three measure, $s$, $l$ and $r$ fluctuate) for the chimera states is relatively smaller (cf. Fig.~2 in the main text). We emphasize that such a transition emerges for a range of $\alpha$ values when other parameters will also change.  
\begin{figure}[h!]
	\centering
	\centering
	\includegraphics[width = 9cm,height = 4cm]{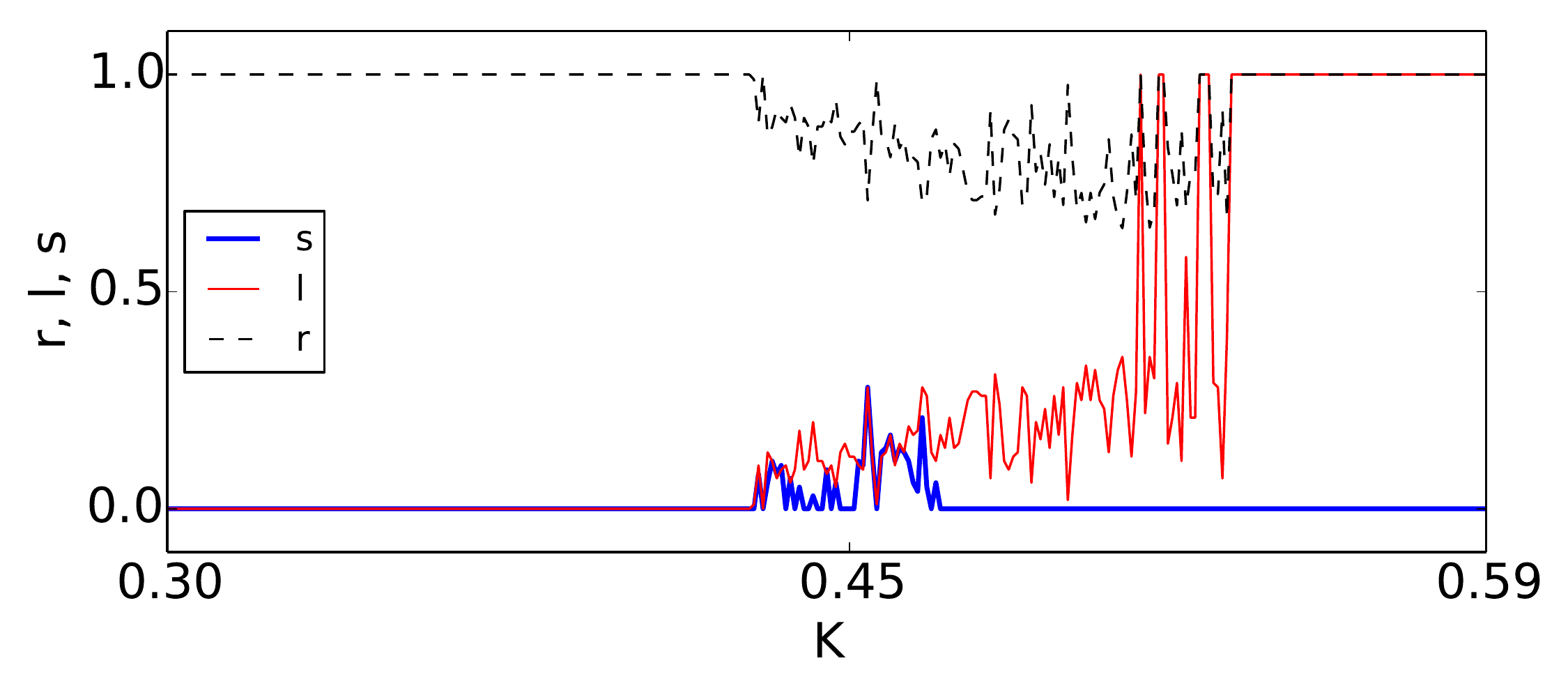}
	\label{}	
	\centering	
	\caption{(Color online) Plots of $r$ (black dashed line), $l$ (Red line) and $s$ (bold blue line) with coupling strength $K$. $\alpha=0.25$, $\Omega_{rf}=0.8$, $I_{rf} = 0.6$.}
	\label{FIG.1}
\end{figure}